\documentclass[a4paper,fleqn,usenatbib]{mnras}
\usepackage[T1]{fontenc}
\usepackage{ae,aecompl}

\usepackage{graphicx}	% Including figure files
\usepackage{amsmath}	% Advanced maths commands
\usepackage{amssymb}	% Extra maths symbols

\title[A degeneracy in DRW modelling of AGN light curves]{A degeneracy in DRW modelling of AGN light curves}

\author[Szymon Koz{\l}owski]{
Szymon Koz{\l}owski,$^{1}$\thanks{E-mail: simkoz@astrouw.edu.pl}
\\
$^{1}$Warsaw University Observatory, Al. Ujazdowskie 4, PL-00-478 Warszawa, Poland
}

\date{Accepted 2016 April 06. Received 2016 April 05; in original form 2016 March 13}

\pubyear{2016}

\begin{document}
\label{firstpage}
\pagerange{\pageref{firstpage}--\pageref{lastpage}}
\maketitle

\begin{abstract}
Individual light curves of active galactic nuclei (AGNs) are nowadays successfully modelled with the damped random walk (DRW) stochastic process,
characterized by the power exponential covariance matrix of the signal, with the power $\beta=1$. 
By Monte Carlo simulation means, we generate mock AGN light curves described by non-DRW stochastic processes ($0.5\leq\beta\leq 1.5$ and $\beta\neq1$) 
and show they can be successfully and well modelled as a single DRW process, obtaining comparable goodness of fits. 
A good DRW fit, in fact, may not mean that DRW is the true underlying process leading to variability and it cannot be used as a proof for it.
When comparing the input (non-DRW) and measured (DRW) process parameters, 
the recovered time-scale (amplitude) increases (decreases) with the increasing input $\beta$. 
In practice, this means that the recovered DRW parameters may lead to biased (or even non-existing) correlations 
of the variability and physical parameters of AGNs if the true AGN variability is caused by non-DRW stochastic processes. The proper way of 
identifying the processes leading to variability are model-independent structure functions and/or power spectral densities
and then using such information on the covariance matrix of the signal in light curve modelling.
\end{abstract}


%%%%%%%%%%%%%%%%%%%%%%%%%%%%%%%%%%%%%%%%%%%%%%%%%%

%%%%%%%%%%%%%%%%% BODY OF PAPER %%%%%%%%%%%%%%%%%%

\section{Introduction}

The damped random walk (DRW) model is an increasingly successful method of quantifying the variability of active galactic nuclei (AGNs; 
\citealt{2009ApJ...698..895K,2010ApJ...708..927K,2010ApJ...721.1014M,2011ApJ...735...80Z, 2013ApJ...765..106Z}). \cite{2009ApJ...698..895K} 
introduce DRW as an underlying stochastic process leading to AGN variability, also known as the continuous-time first order autoregressive process 
[CAR(1)] or Ornstein-Uhlenbeck process (\citealt{1930PhRv...36..823U}).
The model has two parameters, the time-scale $\tau$ after which the light curve becomes uncorrelated and the modified amplitude $\hat\sigma$ 
(\citealt{2010ApJ...708..927K}) or asymptotic amplitude $SF_\infty$ (\citealt{2010ApJ...721.1014M}). 
These two parameters show correlations with the physical parameters of AGNs, such as the black hole mass, luminosity,
Eddington ratio, and rest-frame wavelength. For example, \cite{2009ApJ...698..895K} report that the time-scale $\tau$ is correlated with the black hole mass and luminosity, while the amplitude is anticorrelated with these parameters. \cite{2010ApJ...721.1014M} study $\sim$9000 AGN from Stripe 82 of the Sloan Digital Sky Survey (SDSS) and find that the time-scale $\tau$ is correlated with the rest-frame wavelength and the black hole mass, and does not depend on redshift or luminosity. The asymptotic variability $SF_\infty$ is anticorrelated with the luminosity, rest-frame wavelength, and the Eddington ratio.
\cite{2016arXiv160405858K} reanalysed the same set of SDSS AGN light curves with the `sub-ensemble' structure function (SF) analysis 
that is model-independent and essentially confirms these correlations, albeit with a minute differences in these relations.
He noticed, however, that the SF power-law slope $\gamma$ steepens from $\beta \equiv 2\gamma \approx 1$ for the fainter AGNs
to about $\beta\approx1.2$ for the brightest AGNs, and is independent of the black hole mass.
Such a change means a departure from DRW that is paralleled by the DRW time-scale increase
obtained from light curve modelling (but a bulk of this is the true correlation with the black hole mass).
 
Can a non-DRW stochastic process be successfully and well modelled as DRW, and return correct variability parameters? 
Or will it rather return biased parameters, for example,
longer time-scales for steeper SFs as in the SF analysis from \cite{2016arXiv160405858K}? 
If the latter is the case, then it may have profound implications 
for the reported correlations of variability with the physical parameters of AGNs.
In this paper, we are interested in the modelling of simulated AGN light curves as the DRW process, that are caused by other than DRW underlying processes,
in order to find answers to the above questions.

In Section~\ref{sec:methods}, we present the methodology of simulations and modelling of the quasar light curves,
while in Section~\ref{sec:discussion}, we discuss our findings. The paper is concluded in Section~\ref{sec:summary}.

\section{Methodology}
\label{sec:methods}

We simulate AGN light curves as a single stochastic process with the power exponential covariance matrix of the signal

\begin{equation}
{\rm cov}(\Delta t) = \sigma_s^2 e^{-\left(\frac{|\Delta t|}{\tau}\right)^\beta},
\end{equation}
\noindent
where $\tau>0$ is the decorrelation time-scale, $\sigma_s^2$ is the signal variance, $\Delta t = t_i-t_j$ is the rest-frame time difference (or time lag) between $i$th and $j$th data points,
 and $0<\beta<2$, where $\beta=1$ corresponds to DRW. To simulate a light curve with $N$ points, first,
the ($N\times N$) covariance matrix of the signal 
\begin{equation}
\tiny
C_{ij} = 
 \begin{pmatrix}
  \sigma_s^2 & \sigma_s^2 e^{-\left(\frac{|t_1-t_2|}{\tau}\right)^\beta} & \cdots & \sigma_s^2 e^{-\left(\frac{|t_1-t_N|}{\tau}\right)^\beta} \\
  \sigma_s^2 e^{-\left(\frac{|t_2-t_1|}{\tau}\right)^\beta} & \sigma_s^2& \cdots & \sigma_s^2 e^{-\left(\frac{|t_2-t_N|}{\tau}\right)^\beta} \\
  \vdots  & \vdots  & \ddots & \vdots  \\
  \sigma_s^2 e^{-\left(\frac{|t_N-t_1|}{\tau}\right)^\beta} & \sigma_s^2 e^{-\left(\frac{|t_N-t_2|}{\tau}\right)^\beta} & \cdots & \sigma_s^2 
 \end{pmatrix} 
\end{equation}
\noindent
must be Cholesky-decomposed as $\mathbf{C} = \mathbf{L}^{\rm T}\mathbf{L}$ (e.g., \citealt{1992nrfa.book.....P}), where $\mathbf{L}$ is the upper triangular matrix. 
Next, the light curve is obtained from $\mathbf{y}=\mathbf{Lr}$, where $\mathbf{r}$ is a vector of Gaussian deviations with the variance of unity (e.g., \citealt{2011ApJ...735...80Z}). Finally, we add the photometric noise $n_i(y_i)$ dependent on the magnitude $y_i$, $y_i = y_i + {\rm G}(n_i(y_i))$, 
that is drawn from a Gaussian (G) distribution of the true SDSS photometric noise (\citealt{2007AJ....134..973I}).

We simulate sets of a 1000 light curves in a range $0.5\leq\beta\leq 1.5$, spaced every 0.05. 
With the exception of the case with $\beta=1$, they are regarded as the non-DRW stochastic processes.
The light curves have the mean magnitude $\langle r\rangle=17$ mag and the noise properties of the SDSS Stripe82 quasars (\citealt{2007AJ....134..973I}). 
The input time-scale is $\tau=500$ days, the asymptotic variability amplitude is $SF_\infty=0.18$ mag, 
the light curve length is 8000 days, and the cadence is 20 days (hence 400 points).
Exemplary simulated light curves for $\beta=0.5$, 1.0, and 1.5, are shown in Fig.~\ref{fig:LCs}.

Subsequently, the light curves are modelled with DRW 
[see appendix in \citealt{2010ApJ...708..927K} for fast (only $O(N)$ operations for a light curve with $N$ points) modelling with DRW, also \citealt{2009ApJ...698..895K,2010ApJ...721.1014M, 2011ApJ...735...80Z, 2013ApJ...765..106Z}], i.e., with fixed $\beta=1$.

\begin{figure}
%\centering
\begin{center}
\includegraphics[width=7.5cm]{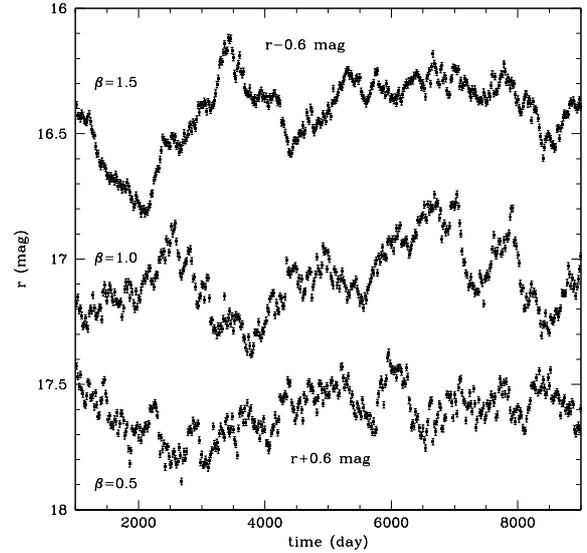}
\end{center}
\caption{Typical simulated AGN light curves with $\langle r\rangle=17$ mag, $\tau=500$ days, $SF_\infty=0.18$ mag, 400 data points and the length of 8000 days, with $\beta=1.5$ (top), $\beta=1.0$ (DRW; middle), and $\beta=0.5$ (bottom).
The top and bottom light curves are shifted by $\pm0.6$ mag for clarity.}
\label{fig:LCs}
\end{figure}

%%%%%%%%%%%%%%%%%%%%%  DISCUSSION  %%%%%%%%%%%%%%%%%%%%

\section{Discussion}
\label{sec:discussion}

\begin{figure}
%\centering
\begin{center}
\includegraphics[width=7.5cm]{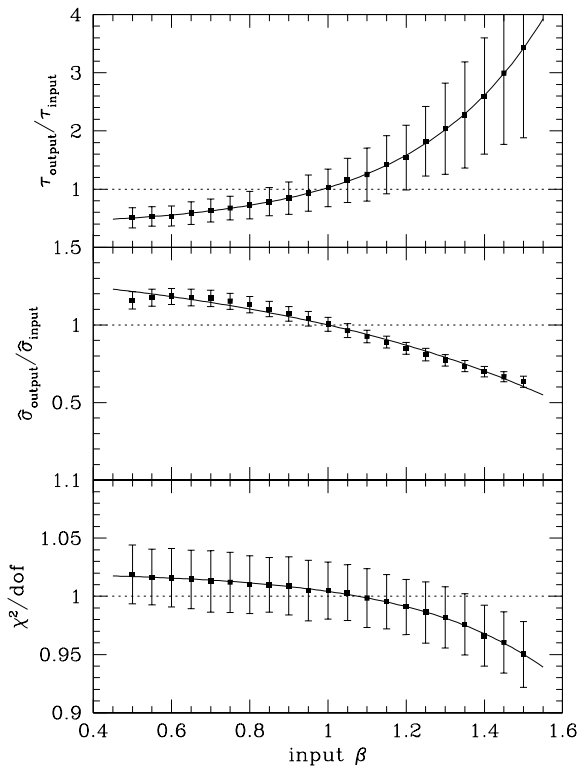}
\end{center}
\caption{Ratios of the median recovered to input parameters (top and middle panels) and $\chi^2$/dof (bottom panel) as a function of the input power $\beta$.
Each point corresponds to a 1000 simulated light curves with $\tau=500$ days, $SF_\infty=0.18$ mag, 400 data points and the length of 8000 days, that are modelled as DRW ($\beta=1$).
Light curves with input $\beta<1$ ($\beta>1$) modelled as DRW, have underestimated time-scales and overestimated amplitudes (overestimated time-scales and underestimated amplitudes) 
while the goodness of fit weakly improves with increasing $\beta$. The error bars are in fact dispersions of the measured values calculated as 0.74 interquartile range (IQR) of the recovered parameter distributions.}
\label{fig:ratios}
\end{figure}

In Fig.~\ref{fig:ratios}, we present the results of modelling AGN light curves as the DRW process, for which the underlying processes were set to be, 
with the exception of $\beta=1$, non-DRW. For each $0.5\leq\beta\leq 1.5$, spaced every 0.05,
we model a 1000 light curves and calculate the median measured parameters along with the dispersions, measured as 0.74 of the interquartile range of these distributions (see \citealt{2012ApJ...753..106M,2016arXiv160405858K}). 
Because the recovered parameters are also a function of the ratio of the time-scale $\tau$ to the experiment length (Koz{\l}owski in preparation), 
we normalize the returned parameters to be unity for $\beta=1$ ($\tau$ is divided by 0.86 and $\hat{\sigma}$ is divided by 0.96).

In top panel of Fig.~\ref{fig:ratios}, we show the ratio of the median of the measured time-scales $\tau$ to the input value ($\tau=500$ days)
as a function of the input parameter $\beta$. We see that the returned time-scale is 
correlated with $\beta$, but also the goodness of fit improves with increasing $\beta$ (bottom panel of Fig.~\ref{fig:ratios}).   
Because in \cite{2016arXiv160405858K} sub-ensemble SFs steepen from $\beta\approx 1.0$ for fainter AGNs to about $\beta \approx 1.2$ for the brightest ones,
it means that DRW should return longer time-scales for the latter sources, even if the true time-scales were identical. 
The middle panel of Fig.~\ref{fig:ratios}, presents the dependence of the recovered modified amplitudes as a function of $\beta$.
It is obvious that the two parameters are anticorrelated.

Finding the exact (parametric) form of these biases is not the goal of this paper, 
because they will depend on the photometric quality and length of data being analysed.
The goal here is simply to make the point about their existence, their possible implications on (mis)understanding of AGN physics,
and to provide a solution to avoid them. Because in \cite{2016arXiv160405858K} $\beta\approx1$ and weakly changes (to $\sim$1.2 for the brightest AGNs), the positive side is that
modest deviations from the DRW model seem to be nearly unimportant for the estimated variability parameters and 
they weakly affect the correlations with the physical AGN parameters in \cite{2010ApJ...721.1014M}. 
The increase of the input (or true) $\beta$ from 1.0 to 1.2 leads to the overestimation of the time-scale $\tau$ by a factor of 1.5 (Fig.~\ref{fig:ratios}).
The negative side is that typical AGN light curves are not good enough to notice the deviations from DRW and so one may misinterpret parameters.
Because SFs or power spectral densities are a model-independent means of estimating the shape of the 
covariance function of the signal (e.g., \citealt{2016arXiv160405858K}), 
one should rather estimate $\beta$ this way, and then use it as input parameter in direct light curve modelling to obtain
correct model parameters. \cite{2013ApJ...765..106Z} discusses modelling of light curves with additional parameters to that from DRW.

%%%%%%%%%%%%%%%%%%%%%  SUMMARY  %%%%%%%%%%%%%%%%%%%%

\section{Conclusions}
\label{sec:summary}

In this paper, we have been interested if AGN variability caused by non-DRW stochastic 
processes can be well modelled with a single DRW process, nowadays frequently considered in AGN variability studies.
By simulation means, we have tested the implications of modelling non-DRW processes on the measured DRW model parameters
and found that they are biased, 
where the time-scale increases and the amplitude decreases with the increasing input parameter $\beta$ (the 
power of the power exponential covariance matrix of the signal). 
Equally important finding here is the goodness of fit being unable to recognize what process is being modelled, hence, the word `degeneracy' in the title.
A good DRW fit should not and cannot be used as a proof for DRW as the true underlying process leading to variability. 
Instead, the covariance matrix of the signal should be obtained from model-independent measures of variability such as 
the structure functions or power spectral densities, and serve as input for the covariance matrix used in direct light curve modelling.

An answer to a question if DRW was a good model describing the AGN variability would be yes. Yes, because both DRW and non-DRW
processes described by the power exponential covariance matrix of the signal are very well modelled by a single DRW process.
And yes, because \cite{2016arXiv160405858K} based on model-independent structure functions shows that $\beta\approx1$ for 9000 SDSS AGNs, consistent with DRW.
The caveat is, however, that some of the underlying processes may be non-DRW, as indicated by \cite{2011ApJ...743L..12M} and 
\cite{2015MNRAS.451.4328K} based on steeper than DRW power spectral 
distributions and structure functions of {\it Kepler} AGNs (the conversion between the power spectral density modelled
as a single power-law with the slope $\alpha$ is $\beta=-0.5\alpha$, so we have explored in this paper $-1<\alpha<-3$,
where $\alpha\approx-3$ was reported by \cite{2011ApJ...743L..12M}).
Then the DRW light curve modelling will not be able to identify a non-DRW process and will return biased DRW variability parameters.

As a matter of fact, in \cite{2010ApJ...708..927K}, we already modelled deterministic processes (non-stochastic) such as periodic variable stars and 
found that DRW modelling provides a good description, where the time-scale $\tau$ is identified with the variability period.

\section*{Acknowledgements}

I am grateful to Chris Kochanek for discussions of this topic.
This work has been supported by the Polish National Science Centre OPUS grant number 2014/15/B/ST9/00093 and MAESTRO grant number 2014/14/A/ST9/00121.

%%%%%%%%%%%%%%%%%%%%%%%%%%%%%%%%%%%%%%%%%%%%%%%%%%



%\bibliographystyle{mnras}


%%%%%%%%%%%%%%%%%%%%%%%%%%%%%%%%%%%%%%%%%%%%%%%%%%

% Don't change these lines
\bsp	% typesetting comment
\label{lastpage}
\end{document}